\documentclass[prd,twocolumn,preprintnumbers,showpacs,nofootinbib]{revtex4}
\usepackage{amsmath,amssymb,bm,graphicx}

\renewcommand\d{\partial}
\newcommand\+{\dagger}

\newcommand\x{{\bm{x}}}
\newcommand\p{{\bm{p}}}
\renewcommand\H{\mathcal{H}}

\begin{document}
\preprint{MIT-CTP 4103}

\title{Is a color superconductor topological?}
\author{Yusuke~Nishida}
\affiliation{Center for Theoretical Physics,
             Massachusetts Institute of Technology,
             Cambridge, Massachusetts 02139, USA}

\begin{abstract}
 A fully gapped state of matter, whether insulator or superconductor,
 can be asked if it is topologically trivial or nontrivial.  Here we
 investigate topological properties of superconducting Dirac fermions in
 3D having a color superconductor as an application.  In the chiral
 limit, when the pairing gap is parity even, the right-handed and
 left-handed sectors of the free space Hamiltonian have nontrivial
 topological charges with opposite signs.  Accordingly, a vortex line in
 the superconductor supports localized gapless right-handed and
 left-handed fermions with the dispersion relations $E=\pm vp_z$ ($v$ is
 a parameter dependent velocity) and thus propagating in opposite
 directions along the vortex line.  However, the presence of the fermion
 mass immediately opens up a mass gap for such localized fermions and
 the dispersion relations become $E=\pm v\sqrt{m^2+p_z^2}$.  When the
 pairing gap is parity odd, the situation is qualitatively different.
 The right-handed and left-handed sectors of the free space Hamiltonian
 in the chiral limit have nontrivial topological charges with the same
 sign and therefore the presence of the small fermion mass does not open
 up a mass gap for the fermions localized around the vortex line.  When
 the fermion mass is increased further, there is a topological phase
 transition at $m=\sqrt{\mu^2+\Delta^2}$ and the localized gapless
 fermions disappear.  We also elucidate the existence of gapless surface
 fermions localized at a boundary when two phases with different
 topological charges are connected.  A part of our results is relevant
 to the color superconductivity of quarks.
\end{abstract}

\date{January 2010}
\pacs{21.65.Qr, 12.38.-t}

\maketitle

\section{Introduction}
A fully gapped state of matter can be asked if it is topologically
trivial or nontrivial.  The most well-known topological state of matter
is the quantum Hall effect in two dimensions
(2D)~\cite{Thouless:1982,Kohmoto:1985,Niu:1985}, in which the time
reversal symmetry is explicitly broken.  Recently the time reversal
symmetric extensions of the quantum Hall effect have been theoretically
proposed and experimentally observed both in
2D~\cite{Kane:2005,Bernevig:2006a,Bernevig:2006b}\cite{Konig:2007} and
3D~\cite{Fu:2007,Moore:2007,Roy:2009}\cite{Hsieh:2008,Hsieh:2009,Xia:2009,Zhang:2009,Chen:2009},
which are referred to as topological insulators (the 2D topological
insulator is also known as the quantum spin Hall effect).  Generally the
topological state of matter is characterized by the nontrivial
topological charge of the single-particle Hamiltonian and accompanied by
topologically protected gapless edge/surface states with linear
dispersions (Dirac fermions).

Another class of topological states of matter arises in
superconductors.  Although the ordinary nonrelativistic $s$-wave
superconductor is topologically trivial, the weakly paired phase of the
$p_x+ip_y$ superconductor in 2D is topologically
nontrivial~\cite{Read:1999fn} and Sr$_2$RuO$_4$ is its candidate
material~\cite{Maeno:1994}.  In addition to the gapless edge/surface
states, the vortex in the topological superconductor supports gapless
fermions localized around the vortex core.  It is also known from the
pioneering work by Jackiw and Rossi that the relativistic $s$-wave
superconductor in 2D has the similar properties~\cite{Jackiw:1981ee}.
Such a system has recently received renewed interest because it can be
realized on the surface of the 3D topological insulator in contact with
the $s$-wave superconductor~\cite{Fu:2008}.  It is pointed out that the
Balian-Werthamer state realized in the B phase of the superfluid $^3$He
is also topological~\cite{Schnyder:2008,Roy:2008,Qi:2009a,Volovik:2003}.

Such progress on the discoveries of the topological insulators and
superconductors motivates us to ask the following question: Is a color
superconductor topological?  In order to shed light on this question, we
investigate in this paper the topological properties of superconducting
Dirac fermions in 3D.
In Sec.~\ref{sec:CSC}, we start with the mean-field model Hamiltonian of
the color superconductivity of quarks and point it out that the free
space Hamiltonian can be characterized by a $\mathbb{Z}$ valued
topological charge.  In Sec.~\ref{sec:even_parity}, we compute the
topological charge of the free space Hamiltonian when the pairing gap is
parity even as a function of the fermion mass and show that its value is
closely linked to the existence of gapless fermions localized around a
vortex line.  The low-energy spectrum of such fermions is also
determined in this section.

In Sec.~\ref{sec:odd_parity}, we turn to the case where the pairing gap
is parity odd.  By studying the topological charge of the free space
Hamiltonian, we show that there is a topological phase transition as a
function of the fermion mass, which is also reflected in the existence
of gapless fermions localized around a vortex line.  In
Sec.~\ref{sec:boundary}, we elucidate the existence of gapless surface
fermions localized at a boundary when two phases with different
topological charges are connected.  Finally, Sec.~\ref{sec:summary} is
devoted to the summary of this paper and implications of our results for
the color superconductor are discussed.

\section{Preparations \label{sec:CSC}}
\subsection{Model Hamiltonian for the color superconductor}
We start with the following mean-field model Hamiltonian for the color
superconductivity of quarks~\cite{Alford:2007xm}:
\begin{equation}\label{eq:CSC}
 \begin{split}
  & H_\mathrm{CSC} = \int\!d\x
  \left[\psi_{a,f}^\+\left(-i\bm\alpha\cdot\bm\d+\beta m-\mu\right)
  \delta_{ab}\delta_{fg}\,\psi_{b,g}\right. \\
  &\left. + \frac12\psi_{a,f}^\+\Delta_{ab,fg}(\x)C\gamma^5\psi_{b,g}^*
  - \frac12\psi_{a,f}^T\Delta_{ab,fg}^\+(\x)C\gamma^5\psi_{b,g}\right].
 \end{split}
\end{equation}
Here $C\equiv i\gamma^2\gamma^0$ is the charge conjugation matrix and we
assumed the same mass $m$ and the same chemical potential $\mu$ for all
three colors ($a,b$) and all three flavors ($f,g$).  $\Delta_{ab,fg}$ is
the pairing gap in the Lorentz singlet and even parity 
($J^P=0^+$) channel and its color and flavor structure is specified by
\begin{equation}
 \Delta_{ab,fg}(\x) = \sum_{i=1,2,3}\Delta_i(\x)\epsilon_{iab}\epsilon_{ifg}.
\end{equation}
$\Delta_1=\Delta_2=\Delta_3\neq0$ corresponds to the fully gapped
color-flavor-locked (CFL) phase and
$\Delta_1=\Delta_2=0,\,\Delta_3\neq0$ corresponds to the two flavor
pairing phase where only four quarks are gapped.  In either case, an
appropriate transformation of $\psi_{a,f}$ by a real and orthogonal
matrix in the color and flavor space can bring the Hamiltonian
(\ref{eq:CSC}) into the decoupled form $H_\mathrm{CSC}=\sum_{j=1}^9H_j$,
where
\begin{align}\label{eq:single_quark}
 H_j &= \int\!d\x
 \left[\psi_j^\+\left(-i\bm\alpha\cdot\bm\d
 +\beta m-\mu\right)\psi_j\right. \notag\\
 &\quad \left. + \frac12\psi_j^\+\Delta_j(\x)C\gamma^5\psi_j^*
 - \frac12\psi_j^T\Delta_j^*(\x)C\gamma^5\psi_j\right] \notag\\
 &= \frac12\int\!d\x
 \begin{pmatrix}
  \psi_j^\+ & -\psi_j^TC\gamma^5
 \end{pmatrix} \notag\\
 &\quad
 \begin{pmatrix}
  -i\bm\alpha\cdot\bm\d+\beta m-\mu & \Delta_j(\x) \\
  \Delta_j^*(\x) & i\bm\alpha\cdot\bm\d-\beta m+\mu
 \end{pmatrix}
 \begin{pmatrix}
  \psi_j \\ C\gamma^5\psi_j^*
 \end{pmatrix} \notag\\
 & \equiv \frac12\int\!d\x\,\Psi_j^\+\H_j\Psi_j.
\end{align}
For example, in the CFL phase, we have
$\Delta_{1,2,3}=-\Delta_{4,\dots,8}=\frac12\Delta_9$.  Below we
concentrate on one sector with the nonzero gap $\Delta_j\neq0$ and
suppress the index of $j$.

The single-particle Hamiltonian $\H$ in the Nambu--Gor'kov representation
has the charge conjugation symmetry:
\begin{equation}
 \mathcal{C}^{-1}\H\mathcal{C} = -\H^*
  \qquad\text{with}\qquad
  \mathcal{C} \equiv
  \begin{pmatrix}
   0 & -C\gamma^5 \\
   C\gamma^5 & 0
  \end{pmatrix}.
\end{equation}
Because $\H\Phi=E\Phi$ leads to
$\H(\mathcal{C}\Phi^*)=-E(\mathcal{C}\Phi^*)$, the spectrum is
symmetric under $E\leftrightarrow-E$.  Furthermore, when the phase of
$\Delta(\x)$ is uniform over the space (i.e. no supercurrent), one can
choose $\Delta(\x)$ to be a real function.\footnote{If we did not make
this choice, we need to modify the definition of $\mathcal{T}$.}  Then
$\H$ has the time
reversal symmetry:
\begin{equation}
 \mathcal{T}^{-1}\H\mathcal{T} = \H^*
  \qquad\text{with}\qquad
  \mathcal{T} \equiv
  \begin{pmatrix}
   \gamma^1\gamma^3 & 0 \\
   0 & \gamma^1\gamma^3
  \end{pmatrix}.
\end{equation}
Because $\H\Phi=E\Phi$ leads to
$\H(\mathcal{T}\Phi^*)=E(\mathcal{T}\Phi^*)$ and
$\Phi^\+\mathcal{T}\Phi^*=0$, the spectrum is at least doubly
degenerate.  Therefore, we find that our Hamiltonian $\H$ belongs to the
symmetry class DIII in the terminology of
Refs.~\cite{Zirnbauer:1996,Altland:1997}.\footnote{Class DIII
Hamiltonians are meant to have the charge conjugation and time reversal
symmetries with properties $\mathcal{C}^T=\mathcal{C}$ and
$\mathcal{T}^T=-\mathcal{T}$~\cite{Schnyder:2008}.}

In Sec.~\ref{sec:odd_parity}, we will also consider the case where the
pairing takes place in the odd parity channel ($J^P=0^-$).  Such a case
is described by simply replacing $\Delta(\x)$ in Eqs.~(\ref{eq:CSC}) and
(\ref{eq:single_quark}) with $\gamma^5\Delta(\x)$.  The resulting
Hamiltonian has the same charge conjugation and time reversal
symmetries and thus belongs to the symmetry class DIII again.

\subsection{Topological charge for class DIII Hamiltonians in 3D}
According to Ref.~\cite{Schnyder:2008}, fully gapped 3D Hamiltonians
belonging to the symmetry class DIII can be classified by a $\mathbb{Z}$
valued topological charge.  The topological charge is defined for the
free space Hamiltonian where $\Delta(\x)=\Delta^*(\x)=\Delta_0$ is a
constant.  Suppose the single-particle Hamiltonian in the momentum space
$\H_\p\equiv e^{-i\p\cdot\x}\,\H\,e^{i\p\cdot\x}$ is diagonalized as
\begin{equation}
 \H_\p = U_\p
  \begin{pmatrix}
   D_\p & 0 \\ 0 & -D_\p
  \end{pmatrix}
   U_\p^\+,
\end{equation}
where $U_\p$ is a unitary matrix and $D_\p$ is a diagonal matrix with
positive elements.  For the fully gapped Hamiltonian, we can
adiabatically deform $D_\p$ into the identity matrix, which continuously
deforms the Hamiltonian into a ``new Hamiltonian'' $Q_\p$ defined by
\begin{equation}
 Q_\p \equiv U_\p
  \begin{pmatrix}
   \openone & 0 \\ 0 & -\openone
  \end{pmatrix}
   U_\p^\+
\end{equation}
with properties $Q_\p=Q_\p^\+$ and $Q_\p^2=1$.  Because of the charge
conjugation and time reversal symmetries, a unitary transformation can
bring the Hermitian matrix $Q_\p$ into a block off-diagonal
form~\cite{Qi:2009b}:
\begin{equation}\label{eq:q_p}
 Q_\p \to
  \begin{pmatrix}
   0 & q_\p \\
   q_\p^\+ & 0
  \end{pmatrix}.
\end{equation}
$Q_\p^2=1$ leads to $q_\p q_\p^\+=1$ and thus $q_\p$ is a unitary matrix
[U(4) in our case].  The topological charge is provided by the winding
number of $q_\p$ associated with the homotopy group
$\pi_3[U(n\geq2)]=\mathbb{Z}$:
\begin{equation}\label{eq:winding}
 N \equiv \frac1{24\pi^2}\int\!d\p\,\epsilon^{ijk}\,\mathrm{Tr}\!
  \left[(q_\p^{-1}\d_iq_\p)(q_\p^{-1}\d_jq_\p)(q_\p^{-1}\d_kq_\p)\right].
\end{equation}
When a given Hamiltonian has a nonzero topological charge, such a system
is said to be {\em topological\/}.  Because Hamiltonians having
different topological charges can not be continuously deformed into each
other without closing energy gaps in their spectrum, the topological
charge defined in Eq.~(\ref{eq:winding}) classifies 3D Hamiltonians
belonging to the symmetry class DIII~\cite{Schnyder:2008}.

It is known in the case of the $p_x+ip_y$ superconductor in 2D that a
nontrivial topological charge of the free space Hamiltonian has a close
connection to the existence of a localized zero energy state in the
presence of a vortex~\cite{Read:1999fn,Gurarie:2007}.  We will see in
the subsequent sections that the same correspondence is true in our
Hamiltonian (\ref{eq:single_quark}) describing superconducting Dirac
fermions in 3D.  We shall work in the chiral representation:
\begin{equation}
 \bm\alpha = \gamma^0\bm\gamma =
  \begin{pmatrix}
   \bm\sigma & 0 \\
   0 &-\bm\sigma
  \end{pmatrix},\qquad
  \beta = \gamma^0 = 
  \begin{pmatrix}
   0 & \openone \\
   \openone & 0
  \end{pmatrix},
\end{equation}
and
\begin{equation}
  \gamma^5 =
  \begin{pmatrix}
   \openone & 0 \\
   0 & -\openone
  \end{pmatrix}. 
\end{equation}

\section{Even parity pairing \label{sec:even_parity}}
We first consider the case where the pairing takes place in the even
parity channel, which is relevant to the color superconductivity of
quarks~\cite{Alford:1997zt,Rapp:1997zu}.

\subsection{Topological charge of a free space Hamiltonian}
When the pairing gap is a constant, we can choose it to be real;
$\Delta(\x)=\Delta^*(\x)=\Delta_0$.  From Eq.~(\ref{eq:single_quark}),
the free space Hamiltonian in the momentum space is given by
\begin{equation}\label{eq:H_even}
 \H_\p =
  \begin{pmatrix}
   \bm\alpha\cdot\bm\p+\beta m-\mu & \Delta_0 \\
   \Delta_0 & -\bm\alpha\cdot\bm\p-\beta m+\mu
  \end{pmatrix}.
\end{equation}
Its energy eigenvalues have the usual form
\begin{equation}
 E_\p = \pm \sqrt{\left(\sqrt{m^2+\p^2}\pm\mu\right)^2+\Delta_0^2}
\end{equation}
and each of them are doubly degenerate (signs are not correlated).  Note
that the spectrum is fully gapped as long as $\Delta_0\neq0$.  The
computation of its topological charge is lengthy but
straightforward.\footnote{Note that the topological charge is invariant
as long as the spectrum is fully gapped.  Therefore one can set
$\mu\to0$ to reduce the computational complication.}  In order to
elucidate the effect of the fermion mass, we shall present the results
for the $m=0$ case and the $m\neq0$ case separately.

\subsubsection{Chiral limit $m=0$}
In the chiral limit $m=0$, because the right-handed sector and the
left-handed sector of the Hamiltonian (\ref{eq:H_even}) are decoupled,
the unitary matrix $q_\p$ in Eq.~(\ref{eq:q_p}) has the block diagonal
form:
\begin{equation}
 q_\p =
  \begin{pmatrix}
   q_{R\p} & 0 \\
   0 & q_{L\p}
  \end{pmatrix}.
\end{equation}
Accordingly, we can define the topological charges for the right-handed
sector and for the left-handed sector independently.  The results
are\,\footnote{The topological charge for the full Hamiltonian
(\ref{eq:CSC}) can be obtained by simply 
summing contributions from all gapped sectors.}
\begin{align}\label{eq:N_R}
 N_R &\equiv \frac1{24\pi^2}\int\!d\p\,\epsilon^{ijk}\,\mathrm{Tr}\!
 \left[(q_{R\p}^{-1}\d_iq_{R\p})(q_{R\p}^{-1}\d_jq_{R\p})
 (q_{R\p}^{-1}\d_kq_{R\p})\right] \notag\\
 &= \frac{\Delta_0}{2|\Delta_0|}
\end{align}
and
\begin{align}\label{eq:N_L}
 N_L &\equiv \frac1{24\pi^2}\int\!d\p\,\epsilon^{ijk}\,\mathrm{Tr}\!
 \left[(q_{L\p}^{-1}\d_iq_{L\p})(q_{L\p}^{-1}\d_jq_{L\p})
 (q_{L\p}^{-1}\d_kq_{L\p})\right] \notag\\
 &= -\frac{\Delta_0}{2|\Delta_0|}.
\end{align}
We find that each sector is topologically nontrivial having the nonzero
topological charge.\footnote{The half-integer value of $N$ is common to
relativistic fermions because $q_\p$ is noncompact at $|\p|\to\infty$.
See, e.g., Ref.~\cite{Schnyder:2008}.}  This implies the existence of a
localized zero energy state for each sector in the presence of a vortex.
However, their signs are opposite and the total topological charge of
the Hamiltonian (\ref{eq:H_even}) is vanishing; $N=N_R+N_L=0$.

\subsubsection{Nonzero fermion mass $m\neq0$}
When the fermion mass is nonzero $m\neq0$, the right-handed and
left-handed sectors are coupled and thus the only total topological
charge $N$ is well defined.  Because the spectrum is fully gapped for
$\Delta_0\neq0$, the inclusion of the fermion mass can not change the
topological charge.  Therefore we find
\begin{equation}\label{eq:N_even2}
 N = 0
\end{equation}
for arbitrary $m$, which means that the system is topologically
trivial.  In the following subsection, we will see how these
observations in the free space are reflected in the spectrum of fermions
localized around a vortex line.

\begin{figure*}[tp]
 \includegraphics[width=0.65\textwidth,clip]{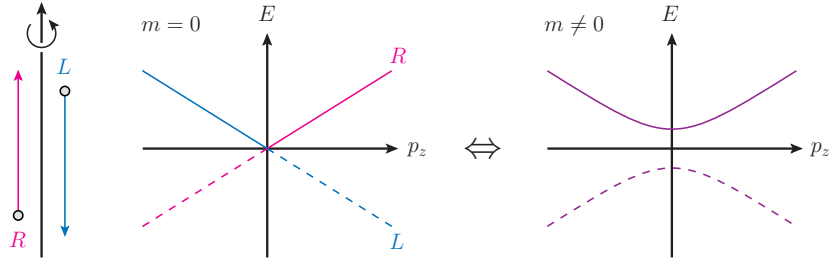}
 \caption{Low-energy spectrum of fermions localized around the vortex
 line in the chiral limit (left panel) and with the nonzero fermion mass
 for the even parity pairing (right panel).  The $E<0$ part of the
 spectrum is redundant in the superconductor and thus shown by the
 dashed line. \label{fig:spectrum}}
\end{figure*}

\subsection{Spectrum of fermions localized around a vortex line
  \label{sec:spectrum_even}}
The spectrum of fermions in the presence of a singly quantized vortex
line is obtained by solving the Bogoliubov--de Genne equation:
\begin{equation}\label{eq:BdG_even}
 \begin{pmatrix}
  -i\bm\alpha\cdot\bm\d+\beta m-\mu & e^{i\theta}|\Delta(r)| \\
  e^{-i\theta}|\Delta(r)| & i\bm\alpha\cdot\bm\d-\beta m+\mu
 \end{pmatrix}
 \Phi(\x) = E \Phi(\x).
\end{equation}
Here we assumed that the vortex line extends in the $z$ direction and
$\Delta(\x)$ does not depend on $z$;
$\Delta(\x)=e^{i\theta}|\Delta(r)|$ where $(r,\theta,z)$ are
cylindrical polar coordinates.
Note that we do not make any assumptions on the form of $|\Delta(r)|$
except that it has a nonvanishing asymptotic value;
$|\Delta(r\to\infty)|>0$.  Therefore the existence of localized fermions
that we will find below is independent of the vortex profile and thus in
this sense they are {\em universal\/}.  This would be because these
solutions have topological origins and, in particular, the zero energy
solutions are guaranteed by the index
theorem~\cite{Jackiw:1981ee,Weinberg:1981eu}.  In contrast, there will
be other Caroli--de Gennes--Matricon-type bound fermions on the vortex
line which typically have the energy gap
$\sim|\Delta(\infty)|^2/\mu$~\cite{Caroli:1964}.  Because their spectrum
depends on the vortex profile, we shall not investigate such
nonuniversal solutions in this paper.

Because of the translational invariance in the $z$ direction, we look
for solutions of the form
\begin{equation}
 \Phi(r,\theta,z) = e^{ip_zz}\phi_{p_z}\!(r,\theta).
\end{equation}
We rewrite the Hamiltonian in Eq.~(\ref{eq:BdG_even}) as
\begin{equation}
 \begin{split}
  e^{-ip_zz}\,\H\,e^{ip_zz}
  &= \left.\H\right|_{m=p_z=0} +
  \begin{pmatrix}
   \alpha_zp_z+\beta m & 0 \\
   0 & -\alpha_zp_z-\beta m
  \end{pmatrix} \\
  &\equiv \H_0 + \delta\H.
 \end{split}
\end{equation}
We first construct zero energy solutions for $\H_0$ at $m=p_z=0$ and
then determine their dispersion relations with treating $\delta\H$
($m,p_z\neq0$) as a perturbation.

\subsubsection{Zero energy solutions at $m=p_z=0$}
Consider the zero energy Bogoliubov--de Genne equation at $m=p_z=0$;
$\H_0\phi_0=0$.  We can find two exponentially localized solutions (see
Appendix~\ref{sec:derivation_even}):
\begin{equation}\label{eq:zero-energy_even1}
 \phi_R \equiv \frac{e^{i\frac\pi4}}{\sqrt\lambda}
 \begin{pmatrix}
  J_0(\mu r) \\
  ie^{i\theta}J_1(\mu r) \\
  0 \\ 0 \\
  e^{-i\theta}J_1(\mu r) \\
  -iJ_0(\mu r) \\
  0 \\ 0
 \end{pmatrix}
 e^{-\int_0^r|\Delta(r')|dr'}
\end{equation}
and
\begin{equation}\label{eq:zero-energy_even2}
 \phi_L \equiv \frac{e^{-i\frac\pi4}}{\sqrt\lambda}
  \begin{pmatrix}
   0 \\ 0 \\
   J_0(\mu r) \\
   -ie^{i\theta}J_1(\mu r) \\
   0 \\ 0 \\
   e^{-i\theta}J_1(\mu r) \\
   iJ_0(\mu r)
  \end{pmatrix}
  e^{-\int_0^r|\Delta(r')|dr'},
\end{equation}
where $\lambda$ is a normalization constant:
\begin{equation}
 \lambda = 2\pi\int_0^\infty\!dr\,r
  \left[2J_0^2(\mu r)+2J_1^2(\mu r)\right]e^{-2\int_0^r|\Delta(r')|dr'}.
\end{equation}
These two solutions have definite chirality;
$\gamma^5\phi_{R/L}=\pm\phi_{R/L}$, and hence their index.

\subsubsection{Perturbations in terms of $m$ and $p_z$}
\begin{figure}[tp]
 \includegraphics[width=0.4\textwidth,clip]{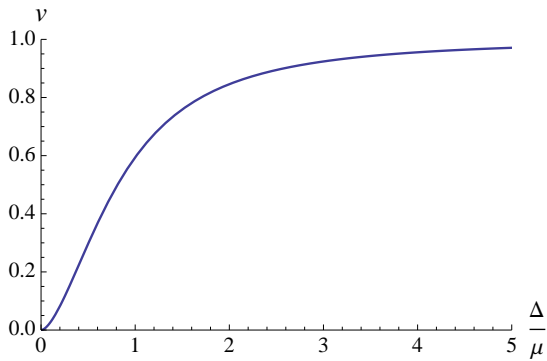}
 \caption{The velocity $v$ in Eq.~(\ref{eq:velocity_even2}) as a
 function of $\Delta/\mu$. \label{fig:velocity}}
\end{figure}

We now evaluate matrix elements of $\delta\H$ with respect to $\phi_R$
and $\phi_L$.  It is easy to find
\begin{equation}
 \int_0^{2\pi}d\theta\int_0^\infty\!dr\,r
  \begin{pmatrix}
   \phi_R^\+\,\delta\H\,\phi_R
   & \phi_R^\+\,\delta\H\,\phi_L \\
   \phi_L^\+\,\delta\H\,\phi_R
   & \phi_L^\+\,\delta\H\,\phi_L
  \end{pmatrix} \\
 = v
  \begin{pmatrix}
   p_z & -im \\ im & -p_z
  \end{pmatrix},
\end{equation}
where we defined the parameter dependent velocity $|v|\leq1$ in units of
the speed of light:
\begin{equation}\label{eq:velocity_even}
 v \equiv
  \frac{\int_0^\infty\!dr\,r\left[J_0^2(\mu r)-J_1^2(\mu r)\right]
  e^{-2\int_0^r|\Delta(r')|dr'}}
  {\int_0^\infty\!dr\,r\left[J_0^2(\mu r)+J_1^2(\mu r)\right]
  e^{-2\int_0^r|\Delta(r')|dr'}}.
\end{equation}
Therefore, when $m=0$, the right-handed and left-handed fermions
localized around the vortex line have the gapless dispersion relations:
\begin{equation}
 E = v p_z \qquad\text{and}\qquad E = -v p_z,
\end{equation}
respectively.  They have opposite velocities and thus propagate in
opposite directions along the vortex line (Fig.~\ref{fig:spectrum}).  In
the simple case where $|\Delta(r)|=\Delta>0$ is a constant, the velocity
$v$ in Eq.~(\ref{eq:velocity_even}) can be evaluated as
\begin{equation}\label{eq:velocity_even2}
 \begin{split}
  v &= \frac{\mu^2}{\mu^2+\Delta^2}\frac{\mathcal E(-\mu^2/\Delta^2)}
  {\mathcal E(-\mu^2/\Delta^2)-\mathcal K(-\mu^2/\Delta^2)} - 1 \\
  &\to
  \begin{cases}
   \left(\ln\frac{4\mu}{\Delta}-1\right)\left(\frac{\Delta}{\mu}\right)^2
   + \cdots & \left(\frac{\Delta}{\mu} \to 0\right) \medskip\\
   1 - \frac{3}{4}\left(\frac{\mu}{\Delta}\right)^2 + \cdots
   & \left(\frac{\Delta}{\mu} \to \infty\right),
  \end{cases}
 \end{split}
\end{equation}
which is plotted in Fig.~\ref{fig:velocity} as a function of
$\Delta/\mu$.  Here $\mathcal K\,(\mathcal E)$ is the complete elliptic
integral of the first (second) kind.

On the other hand, when $m\neq0$, the right-handed and left-handed
fermions are mixed and their spectrum exhibits the mass gap provided by
$vm$ (Fig.~\ref{fig:spectrum}):
\begin{equation}
 E = \pm v \sqrt{m^2+p_z^2}.
\end{equation}
This is closely linked to our previous observations on the topological
charge of the free space Hamiltonian.  In the chiral limit $m=0$, each
of the right-handed and left-handed sectors is topologically nontrivial
($N_R=-N_L\neq0$) and thus the vortex line supports the localized
gapless fermions.  However, once the fermion mass $m\neq0$ is
introduced, the total Hamiltonian is topologically trivial ($N=0$) and
thus the vortex line no longer supports the gapless fermions.
Nevertheless, as long as $vm\ll\Delta(r\to\infty)$ is satisfied, the
mass gap of such localized fermions is much smaller than the energy gap
of bulk fermions and thus they can be important low-energy degrees of
freedom.

\subsubsection{Nonperturbative solutions at $\mu=0$}
The above results rely on the perturbations in terms of $m$ and $p_z$.
In the special case where $\mu=0$, we can obtain the exact dispersion
relations of the localized fermions and their eigenfunctions with
arbitrary $m$ and $p_z$ (see Appendix~\ref{sec:derivation_even}).  The
two solutions to the Bogoliubov--de Genne equation (\ref{eq:BdG_even})
at $\mu=0$ are found to be
\begin{equation}\label{eq:nonzero_even}
 \Phi_\pm(r,\theta,z) = \frac{e^{-i\frac\pi4}}{\sqrt{\lambda_\pm'}}
  \begin{pmatrix}
   p_z+E_\pm \\ 0 \\ m \\ 0 \\
   0 \\ -i(p_z+E_\pm) \\ 0 \\ im
  \end{pmatrix}
  e^{ip_zz-\int_0^r|\Delta(r')|dr'}
\end{equation}
with $E_\pm=\pm\sqrt{m^2+p_z^2}$.  The discussions given above are valid
in this special case too.

\subsection{Effective 1D Hamiltonian along a vortex line}
Because the bulk fermions are gapped, the low-energy effective
Hamiltonian of the system in the chiral limit $m=0$ should involve the
gapless fermions existing along the vortex line.  In order to write down
the effective 1D Hamiltonian, we expand the fermion operator $\Psi$ in
Eq.~(\ref{eq:single_quark}) in terms of the eigenfunctions of $\H$:
\begin{equation}
 \Psi(\x) = \int\!\frac{dp_z}{2\pi}
  \left(a_{p_z}e^{ip_zz}\phi_R+b_{p_z}e^{ip_zz}\phi_L+\cdots\right),
\end{equation}
where $a_{p_z}$ and $b_{p_z}$ are quasiparticle operators associated
with the gapless right-handed and left-handed fermions, respectively.
Because of the pseudoreality condition $\Psi=\mathcal{C}\Psi^*$ and the
property of the solutions $\mathcal{C}\phi_{R/L}^*=\phi_{R/L}$, we have
$a_{p_z}^\dagger=a_{-p_z}$ and $b_{p_z}^\dagger=b_{-p_z}$.  The
quasiparticle operators obeying such conditions are called as
{\em Majorana fermions\/} in condensed matter
literatures~\cite{Read:1999fn,Fu:2008,Schnyder:2008,Roy:2008,Qi:2009a}.
From the Hamiltonian $H=\frac12\int\!d\x\,\Psi^\+\H\Psi$, the effective
1D Hamiltonian becomes
\begin{equation}
 H_\mathrm{1D} = \frac{v}2\int\!\frac{dp_z}{2\pi}
  \left(p_za_{p_z}^\+a_{p_z}-p_zb_{p_z}^\+b_{p_z}\right).
\end{equation}
When the nonzero fermion mass $m\neq0$ is present, there are additional
terms mixing $a_{p_z}$ and $b_{p_z}$:
\begin{equation}
 \begin{split}
  H_\mathrm{1D} = \frac{v}2\int\!\frac{dp_z}{2\pi}
  & \left(p_z a_{p_z}^\+a_{p_z}-p_z b_{p_z}^\+b_{p_z}\right.\\
  & \left.+imb_{p_z}^\+a_{p_z}-ima_{p_z}^\+b_{p_z}\right).
 \end{split}
\end{equation}

\section{Odd parity pairing \label{sec:odd_parity}}
So far we have discussed the case where the pairing takes place in the
even parity channel.  The situation is qualitatively different when the
pairing takes place in the odd parity channel
[$\Delta(\x)\to\gamma^5\Delta(\x)$].

\subsection{Topological charge of a free space Hamiltonian}
When the pairing gap is a constant $\Delta(\x)=\Delta^*(\x)=\Delta_0$,
the free space Hamiltonian in the momentum space is given by
\begin{equation}\label{eq:H_odd}
 \H_\p =
  \begin{pmatrix}
   \bm\alpha\cdot\bm\p+\beta m-\mu & \gamma^5\Delta_0 \\
   \gamma^5\Delta_0 & -\bm\alpha\cdot\bm\p-\beta m+\mu
  \end{pmatrix}.
\end{equation}
Its energy eigenvalues have the form
\begin{equation}
 E_\p = \pm \sqrt{m^2+\p^2+\mu^2+\Delta_0^2 \pm
  2\sqrt{m^2\left(\mu^2+\Delta_0^2\right)+\p^2\mu^2}}
\end{equation}
and each of them are doubly degenerate (signs are not correlated).  Note
that even if $\Delta_0\neq0$, the energy gap in the spectrum is closed
at $m^2=\mu^2+\Delta_0^2$.  This fact will have important consequences
on the topological charge of the free space Hamiltonian and thus on the
existence of localized zero energy states in the presence of a vortex.

\subsubsection{Chiral limit $m=0$}
In the chiral limit $m=0$, we can define the topological charges for
the right-handed sector and for the left-handed sector independently.
Compared to the even parity pairing, $\gamma^5$ in front of $\Delta_0$
flips the sign of the pairing gap only in the left-handed sector.
Therefore, from Eqs.~(\ref{eq:N_R}) and (\ref{eq:N_L}), we easily obtain
\begin{equation}\label{eq:N_odd}
 N_R = \frac{\Delta_0}{2|\Delta_0|}
  \qquad\text{and}\qquad
  N_L = \frac{\Delta_0}{2|\Delta_0|}.
\end{equation}
We find that each sector is topologically nontrivial having the nonzero
topological charge.  The striking difference of the odd parity pairing
from the even parity pairing is that their signs are the same and thus
the total topological charge of the Hamiltonian (\ref{eq:H_odd}) is
nonvanishing;
\begin{equation}
 N = N_R + N_L = \frac{\Delta_0}{|\Delta_0|}.
\end{equation}
This implies, unlike the even parity pairing, that the existence of
localized zero energy states in the presence of a vortex is robust
against the inclusion of the small fermion mass.

\subsubsection{Nonzero fermion mass $m\neq0$}
When the fermion mass is nonzero $m\neq0$, the only total topological
charge $N$ is well defined.  Because the spectrum remains gapped as long
as $m^2<\mu^2+\Delta_0^2$, the topological charge remains the same;
$N=\Delta_0/|\Delta_0|$.  However, when the fermion mass exceeds the
critical value $m^2=\mu^2+\Delta_0^2$ at which the energy gap in the
spectrum closes, there is a quantum phase transition to another fully
gapped phase characterized by its vanishing topological
charge\,\footnote{This phase is continuously connected to the Hamiltonian
with $\mu,\Delta_0\to0$ where the topological charge can be computed
most easily.}:
\begin{equation}\label{eq:N_odd2}
 N = 
\begin{cases}\displaystyle
 \frac{\Delta_0}{|\Delta_0|} & \text{for}\ \ m^2<\mu^2+\Delta_0^2 \medskip\\
 \ \ \,0 & \text{for}\ \ m^2>\mu^2+\Delta_0^2.
\end{cases}
\end{equation}
Because these two phases can not be distinguished by symmetries, the
phase transition between them is a {\em topological phase transition\/}.
This is exactly the same type of the topological phase transition
occurring in the 2D $p_x+ip_y$ superconductor at $\mu=0$ as a function
of the chemical potential~\cite{Read:1999fn,Gurarie:2007}.  Other
quantum phase transitions resulting from the momentum space topology are
extensively discussed in Ref.~\cite{Volovik:2006gt}.  In the following
subsection, we will see how these observations in the free space are
reflected in the existence of gapless fermions localized around a vortex
line.

\subsection{Spectrum of fermions localized around a vortex line}
The spectrum of fermions in the presence of a singly quantized vortex
line is obtained by solving the Bogoliubov--de Genne equation:
\begin{equation}\label{eq:BdG_odd}
 \begin{pmatrix}
  -i\bm\alpha\cdot\bm\d+\beta m-\mu & \gamma^5e^{i\theta}|\Delta(r)| \\
  \gamma^5e^{-i\theta}|\Delta(r)| & i\bm\alpha\cdot\bm\d-\beta m+\mu
 \end{pmatrix}
 \Phi(\x) = E \Phi(\x).
\end{equation}
As we stated in Sec.~\ref{sec:spectrum_even}, we do not make any
assumptions on the form of $|\Delta(r)|$ except that it has a
nonvanishing asymptotic value; $|\Delta(r\to\infty)|>0$.  Because of the
translational invariance in the $z$ direction, we look for solutions of
the form
\begin{equation}
 \Phi(r,\theta,z) = e^{ip_zz}\phi_{p_z}\!(r,\theta).
\end{equation}
We rewrite the Hamiltonian in Eq.~(\ref{eq:BdG_odd}) as
\begin{equation}
 \begin{split}
  e^{-ip_zz}\,\H\,e^{ip_zz}
  &= \left.\H\right|_{p_z=0} +
  \begin{pmatrix}
   \alpha_zp_z & 0 \\
   0 & -\alpha_zp_z
  \end{pmatrix} \\
  &\equiv \H_0 + \delta\H.
 \end{split}
\end{equation}
We first construct zero energy solutions for $\H_0$ at $p_z=0$ and then
determine their dispersion relations with treating $\delta\H$
($p_z\neq0$) as a perturbation.

\subsubsection{Zero energy solutions at $p_z=0$}
Consider the zero energy Bogoliubov--de Genne equation at $p_z=0$;
$\H_0\phi_0=0$.  We can find two potentially normalizable solutions (see
Appendix~\ref{sec:derivation_odd}):
\begin{equation}\label{eq:zero-energy_odd1}
 \phi_R \equiv \frac{e^{i\frac\pi4}}{\sqrt\lambda}
  \begin{pmatrix}
   \mu \bar J_0 \\
   i\sqrt{\mu^2-m^2}\,e^{i\theta}\bar J_1 \\
   m \bar J_0 \\ 0 \\
   \sqrt{\mu^2-m^2}\,e^{-i\theta}\bar J_1 \\
   -i\mu \bar J_0 \\ 
   0 \\ -im \bar J_0
  \end{pmatrix}
  e^{-\int_0^r|\Delta(r')|dr'}
\end{equation}
and
\begin{equation}\label{eq:zero-energy_odd2}
 \phi_L \equiv \frac{e^{i\frac\pi4}}{\sqrt\lambda}
  \begin{pmatrix}
   m \bar J_0 \\ 0 \\
   \mu \bar J_0 \\
   -i\sqrt{\mu^2-m^2}\,e^{i\theta}\bar J_1 \\
   0 \\ -im \bar J_0 \\ 
   -\sqrt{\mu^2-m^2}\,e^{-i\theta}\bar J_1 \\
   -i\mu \bar J_0
  \end{pmatrix}
  e^{-\int_0^r|\Delta(r')|dr'},
\end{equation}
where we introduced shorthand notations;
$\bar J_0\equiv J_0(\sqrt{\mu^2-m^2}\,r)$ and
$\bar J_1\equiv J_1(\sqrt{\mu^2-m^2}\,r)$, and $\lambda$ is a
normalization constant:
\begin{equation}
 \begin{split}
  \lambda &= 2\pi\int_0^\infty\!dr\,r
  \left[2(\mu^2+m^2)\bar J_0^2+2(\mu^2-m^2)\bar J_1^2\right] \\
  &\quad \times e^{-2\int_0^r|\Delta(r')|dr'}.
 \end{split}
\end{equation}
These two solutions in the chiral limit $m=0$ have definite chirality;
$\gamma^5\phi_{R/L}=\pm\phi_{R/L}$, and hence their index.  Note that,
although the nonzero fermion mass $m\neq0$ mixes the right-handed and
left-handed fermions, their gaplessness is preserved.

We now examine if the above two solutions are normalizable or not.  When
$m^2<\mu^2$, they are normalizable owing to the exponentially decaying
factor $e^{-\int_0^r|\Delta(r')|dr'}\to e^{-r|\Delta(\infty)|}$ at
$r\to\infty$.  When $m^2>\mu^2$, we note that 
$\bar J_0=I_0(\sqrt{m^2-\mu^2}\,r)$ and
$\bar J_1=iI_1(\sqrt{m^2-\mu^2}\,r)$ exponentially grow as
$\bar J_0,\bar J_1\to e^{\sqrt{m^2-\mu^2}\,r}$.  Nevertheless, as long
as $m^2<\mu^2+|\Delta(\infty)|^2$ is satisfied, the exponentially
decaying factor dominates and the solutions are still normalizable.
However, when the fermion mass exceeds the critical value
$m^2=\mu^2+|\Delta(\infty)|^2$, the two solutions exponentially grow and
thus they are no longer acceptable solutions.  Therefore we find that
the gapless fermions localized around the vortex line exist only when
$m^2<\mu^2+|\Delta(\infty)|^2$ and they disappear when
$m^2>\mu^2+|\Delta(\infty)|^2$.

Of course this is closely linked to our previous observations on the
topological charge of the free space Hamiltonian; it is topologically
nontrivial ($N\neq0$) for $m^2<\mu^2+|\Delta_0|^2$ and trivial ($N=0$)
for $m^2>\mu^2+|\Delta_0|^2$ and there is a topological phase transition
in between.  Also note the striking difference of the odd parity pairing
from the even parity pairing where the presence of the fermion mass
immediately opened up a mass gap for the fermions localized around the
vortex line.

\subsubsection{Perturbation in terms of $p_z$}
We now evaluate matrix elements of $\delta\H$ with respect to
$\phi_R$ and $\phi_L$ for $m^2<\mu^2+|\Delta(\infty)|^2$.  It is easy to
find
\begin{equation}
 \int_0^{2\pi}d\theta\int_0^\infty\!dr\,r
  \begin{pmatrix}
   \phi_R^\+\,\delta\H\,\phi_R
   & \phi_R^\+\,\delta\H\,\phi_L \\
   \phi_L^\+\,\delta\H\,\phi_R
   & \phi_L^\+\,\delta\H\,\phi_L
  \end{pmatrix} \\
 = v
  \begin{pmatrix}
   p_z & 0 \\ 0 & -p_z
  \end{pmatrix},
\end{equation}
where we defined the parameter dependent velocity $|v|\leq1$ in units of
the speed of light:
\begin{equation}\label{eq:velocity_odd}
 v \equiv
  \frac{\int_0^\infty\!dr\,r\left[(\mu^2-m^2)\bar J_0^2
  -(\mu^2-m^2)\bar J_1^2\right] e^{-2\int_0^r|\Delta(r')|dr'}}
  {\int_0^\infty\!dr\,r\left[(\mu^2+m^2)\bar J_0^2
  +(\mu^2-m^2)\bar J_1^2\right] e^{-2\int_0^r|\Delta(r')|dr'}}.
\end{equation}
Therefore the dispersion relations of the gapless fermions localized
around the vortex line are given by
\begin{equation}
 E = v p_z \qquad\text{and}\qquad E = -v p_z.
\end{equation}
They have opposite velocities and thus propagate in opposite directions
along the vortex line.  In the simple case where $|\Delta(r)|=\Delta>0$
is a constant, the velocity $v$ in Eq.~(\ref{eq:velocity_odd}) can be
evaluated as
\begin{equation}\label{eq:velocity_odd2}
 v = \frac{\mu^2\,\bar{\mathcal E}}
  {\left(\mu^2+\Delta^2\right)\bar{\mathcal E}
  -\left(\mu^2+\Delta^2-m^2\right)\bar{\mathcal K}} - 1,
\end{equation}
where $\bar{\mathcal K}\equiv\mathcal K((m^2-\mu^2)/\Delta^2)$ and
$\bar{\mathcal E}\equiv\mathcal E((m^2-\mu^2)/\Delta^2)$.  When $m=0$,
this is identical to Eq.~(\ref{eq:velocity_even2}) and plotted in
Fig.~\ref{fig:velocity} as a function of $\Delta/\mu$.  The velocity
vanishes at $m=\mu$ and changes its sign for $m>\mu$.

\subsubsection{Nonperturbative solutions at $\mu=0$}
The above results rely on the perturbation in terms of $p_z$.  In the
special case where $\mu=0$, we can obtain the exact dispersion relations
of the localized fermions and their eigenfunctions with arbitrary $p_z$
(see Appendix~\ref{sec:derivation_odd}).  The two solutions to the
Bogoliubov--de Genne equation (\ref{eq:BdG_odd}) at $\mu=0$ are found to
be
\begin{equation}\label{eq:nonzero_odd1}
 \Phi(r,\theta,z) = \frac{e^{i\frac\pi4}}{\sqrt{\lambda'}}
  \begin{pmatrix}
   I_0(mr) \\
   0 \\
   0 \\
   ie^{i\theta}I_1(mr) \\
   0 \\
   -iI_0(mr) \\
   e^{-i\theta}I_1(mr) \\
   0
  \end{pmatrix}
  e^{ip_zz-\int_0^r|\Delta(r')|dr'}
\end{equation}
with $E=p_z$ and
\begin{equation}\label{eq:nonzero_odd2}
 \Phi(r,\theta,z) = \frac{e^{i\frac\pi4}}{\sqrt{\lambda'}}
  \begin{pmatrix}
   0 \\
   -ie^{i\theta}I_1(mr) \\
   I_0(mr) \\
   0 \\
   -e^{-i\theta}I_1(mr) \\
   0 \\
   0 \\
   -iI_0(mr)
  \end{pmatrix}
  e^{ip_zz-\int_0^r|\Delta(r')|dr'}
\end{equation}
with $E=-p_z$.  They are normalizable in the $x$-$y$ plane as long as
$m<\Delta(\infty)$.  The discussions given above are valid in this
special case too.

\subsection{Effective 1D Hamiltonian along a vortex line}
Because the bulk fermions are gapped, the low-energy effective
Hamiltonian of the system with $m^2<\mu^2+|\Delta(\infty)|^2$ should
involve the gapless fermions existing along the vortex line.  In order
to write down the effective 1D Hamiltonian, we expand the fermion
operator $\Psi$ in Eq.~(\ref{eq:single_quark}) in terms of the
eigenfunctions of $\mathcal{H}$:
\begin{equation}
 \Psi(\x) = \int\!\frac{dp_z}{2\pi}
  \left(a_{p_z}e^{ip_zz}\phi_R+b_{p_z}e^{ip_zz}\phi_L+\cdots\right),
\end{equation}
where $a_{p_z}$ and $b_{p_z}$ are Majorana quasiparticle operators
obeying $a_{p_z}^\dagger=a_{-p_z}$ and $b_{p_z}^\dagger=b_{-p_z}$ and
associated with the two gapless fermions.  From the Hamiltonian
$H=\frac12\int\!d\x\,\Psi^\+\H\Psi$, the effective 1D Hamiltonian
becomes
\begin{equation}
 H_\mathrm{1D} = \frac{v}2\int\!\frac{dp_z}{2\pi}
  \left(p_za_{p_z}^\+a_{p_z}-p_zb_{p_z}^\+b_{p_z}\right).
\end{equation}

\section{Boundary problems \label{sec:boundary}}
In Secs.~\ref{sec:even_parity} and \ref{sec:odd_parity}, we have
discussed the connection between the nonzero topological charge of the
free space Hamiltonian and the existence of localized gapless fermions
in the presence of a vortex.  Another characteristic of the topological
state of matter is the existence of gapless edge/surface states when it
is terminated by another gapped state having a different topological
charge.  In this section, we will investigate two types of boundary
problems and elucidate the existence of gapless surface fermions
localized at the boundary.

\subsection{Boundary between $\Delta_0>0$ and $\Delta_0<0$
  \label{sec:sign_change}}
We first consider the boundary where the pairing gap
$\Delta(\x)=\Delta(z)$ changes its sign as ($\Delta_\infty>0$):
\begin{equation}\label{eq:boundary}
 \Delta(z) \to
  \begin{cases}
   +\Delta_\infty & \text{for}\ \ z\to\infty \\
   -\Delta_\infty & \text{for}\ \ z\to-\infty.
  \end{cases}
\end{equation}
Because the topological charge of the free space Hamiltonian, when it is
nonzero, depends on the sign of the pairing gap [see
Eqs.~(\ref{eq:N_R}), (\ref{eq:N_L}), and (\ref{eq:N_odd})], the above
boundary connects two gapped phases with different topological charges
and thus we expect the gapless fermions localized at the boundary.  Here
we present the low-energy spectrum of such fermions and their effective
2D Hamiltonian.  Their derivations are analogous to those in the vortex
problems and some details are provided in
Appendix~\ref{sec:derivation_boundary}.

\subsubsection{Chiral limit $m=0$}
In the chiral limit $m=0$, we can find gapless right-handed and
left-handed fermions localized at the boundary.  Their dispersion
relations are given by
\begin{equation}\label{eq:spectrum}
 E = \pm|v|\sqrt{p_x^2+p_y^2},
\end{equation}
where $\p_\perp\equiv(p_x,p_y)$ is the momentum parallel to the boundary
and $|v|\leq1$ is the parameter dependent velocity in units of
the speed of light:
\begin{equation}\label{eq:v_boundary_even}
 v = \frac{\int_{-\infty}^\infty\!dz\,e^{2i\mu z-2\int_0^z\Delta(z')dz'}}
  {\int_{-\infty}^\infty\!dz\,e^{-2\int_0^z\Delta(z')dz'}}.
\end{equation}
In the simple case with $\Delta(z)=\Delta_\infty\,\mathrm{sgn}(z)$, the
velocity can be evaluated as
$v=\Delta_\infty^2/(\Delta_\infty^2+\mu^2)$.
The low-energy effective Hamiltonian involving the two gapless fermions
existing at the 2D boundary becomes
\begin{equation}\label{eq:H_2D}
 H_\mathrm{2D} = \frac{|v|}2\int\!\frac{d\p_\perp}{(2\pi)^2}
  \left[a_{\p_\perp}^\dagger(\bm\sigma_\perp\cdot\p_\perp)a_{\p_\perp}
  - b_{\p_\perp}^\dagger(\bm\sigma_\perp\cdot\p_\perp)b_{\p_\perp}\right]
\end{equation}
with $\bm\sigma_\perp\equiv(\sigma_1,\sigma_2)$.  Here $a_{\p_\perp}$
and $b_{\p_\perp}$ are two-component Majorana quasiparticle operators
obeying $a_{\p_\perp}^\+=a_{-\p_\perp}^T\sigma_1$ and
$b_{\p_\perp}^\+=b_{-\p_\perp}^T\sigma_1$.

\subsubsection{Even parity pairing with $m\neq0$}
As we can expect from the topological charge of the free space
Hamiltonian [Eqs.~(\ref{eq:N_even2}) and (\ref{eq:N_odd2})], the effect
of the nonzero fermion mass $m\neq0$ is qualitatively different for the
even parity pairing and the odd parity pairing.  When the pairing gap is
parity even, the presence of the fermion mass immediately opens up a
mass gap for the fermions localized at the boundary and the dispersion
relations become
\begin{equation}\label{eq:spectrum_even}
 E = \pm|v|\sqrt{m^2+p_x^2+p_y^2}.
\end{equation}
This is linked to the fact that the two phases at both sides
$z\to\pm\infty$ have the same topological charge $N=0$.
Now the effective 2D Hamiltonian has additional terms mixing
$a_{\p_\perp}$ and $b_{\p_\perp}$:
\begin{equation}
 \begin{split}
  H_\mathrm{2D} = \frac{|v|}2\int\!\frac{d\p_\perp}{(2\pi)^2}
  & \left[a_{\p_\perp}^\dagger(\bm\sigma_\perp\cdot\p_\perp)a_{\p_\perp}
  - b_{\p_\perp}^\dagger(\bm\sigma_\perp\cdot\p_\perp)b_{\p_\perp}\right. \\
  & \left.+im b_{\p_\perp}^\+a_{\p_\perp}-ima_{\p_\perp}^\+b_{\p_\perp}\right].
 \end{split}\end{equation}

\subsubsection{Odd parity pairing with $m\neq0$}
On the other hand, when the pairing gap is parity odd, the two phases at
both sides have different topological charges $N=\pm1$ even in the
presence of the small fermion mass.  Accordingly, the boundary still
supports the two localized gapless fermions [Eqs.~(\ref{eq:spectrum})
and (\ref{eq:H_2D})] with the modified velocity $|v|\leq1$:
\begin{equation}\label{eq:v_boundary_odd1}
 v = \frac{\sqrt{\mu^2-m^2}}{\mu}
  \frac{\int_{-\infty}^\infty\!dz\,e^{2i\sqrt{\mu^2-m^2}\,z-2\int_0^z\Delta(z')dz'}}
  {\int_{-\infty}^\infty\!dz\,e^{-2\int_0^z\Delta(z')dz'}}
\end{equation}
for $m^2<\mu^2$ and
\begin{equation}\label{eq:v_boundary_odd2}
 v = \frac{\sqrt{m^2-\mu^2}}{m}
\end{equation}
for $\mu^2<m^2<\mu^2+\Delta_\infty^2$.

When the fermion mass is increased further and exceeds the critical
value $m^2=\mu^2+\Delta_\infty^2$, the resulting solutions are no longer
normalizable [see Eqs.~(\ref{eq:wave_function}) and
(\ref{eq:solutions_odd})] and the gapless surface fermions disappear.
This is linked to the fact that the two phases at both sides
$z\to\pm\infty$ have the same topological charge $N=0$ as a consequence
of the topological phase transition.

\subsection{Boundary between superconductor and vacuum}
We then consider the following boundary that models the interface between
the superconductor at $z<0$ and the chiral symmetry broken vacuum at
$z>0$:
\begin{equation}
  \begin{cases}
   \ \mu,\,\Delta>0,\ m=0 & \text{for}\ \ z<0 \smallskip\\
   \ \mu=\Delta=0,\ m>0 & \text{for}\ \ z>0.
  \end{cases}
\end{equation}
For the even parity pairing, the two gapped phases at both sides have
the same topological charge $N=0$, while they have different topological
charges for the odd parity pairing; $N=1$ at $z<0$ and $N=0$ at $z>0$.

Accordingly, when the pairing gap is parity odd, we can find one gapless
fermion localized at the boundary $z=0$.  The corresponding two zero
energy solutions at $p_x=p_y=0$ are given by
\begin{equation}
 \Phi(\x) \propto
  \begin{pmatrix}
   e^{i\mu z} \\ 0 \\ -ie^{-i\mu z} \\ 0 \\
   ie^{i\mu z} \\ 0 \\ e^{-i\mu z} \\ 0
  \end{pmatrix}e^{\Delta z},\quad
  \begin{pmatrix}
   0 \\ e^{-i\mu z} \\ 0 \\ ie^{i\mu z} \\
   0 \\ -ie^{i\mu z} \\ 0 \\ e^{i\mu z}
  \end{pmatrix}e^{\Delta z}
\end{equation}
for $z<0$ and
\begin{equation}
 \Phi(\x) \propto
  \begin{pmatrix}
   1 \\ 0 \\ -i \\ 0 \\ i \\ 0 \\ 1 \\ 0
  \end{pmatrix}e^{-mz},\quad
  \begin{pmatrix}
   0 \\ 1 \\ 0 \\ i \\ 0 \\ -i \\ 0 \\ 1
  \end{pmatrix}e^{-mz}
\end{equation}
for $z>0$.  
Its effective 2D Hamiltonian becomes
\begin{equation}
 H_\mathrm{2D} = \frac{|v|}2\int\!\frac{d\p_\perp}{(2\pi)^2}
  \left[a_{\p_\perp}^\dagger(\bm\sigma_\perp\cdot\p_\perp)a_{\p_\perp}\right]
\end{equation}
with the velocity
$v=\left(1+\frac{m}{\Delta+i\mu}\right)/\left(1+\frac{m}{\Delta}\right)$
and the Majorana condition $a_{\p_\perp}^\+=a_{-\p_\perp}^T\sigma_1$.
However, when the pairing gap is parity even, we cannot find such
localized zero energy solutions that are continuous at the boundary
$z=0$ and therefore gapless surface fermions do not exist.

\begin{table*}[tp]
 \renewcommand\arraystretch{1.4}
 \caption{Summary of the topological charge of the free space
 Hamiltonian and the low-energy spectrum of fermions localized around a
 vortex line (or a boundary where the pairing gap changes its sign).  $v$
 is a parameter dependent velocity. \label{tab:summary}}
 \begin{tabular}{l|ll|ll}
  \hline\hline
  & \multicolumn{2}{c|}{even parity pairing}
  & \multicolumn{2}{c}{odd parity pairing} \\\cline{2-5}
  & \quad topological charge & \qquad mid-gap state dispersion$\quad$
	  & \quad topological charge & \qquad mid-gap state dispersion \\\hline
  $\ \ m=0\ \ $ & \quad$N_R=-N_L=\frac{\Delta_0}{2|\Delta_0|}$ & \qquad$E=\pm vp_z$
	  & \quad$N_R=N_L=\frac{\Delta_0}{2|\Delta_0|}$ & \qquad$E=\pm vp_z$ \\\hline
  $\ \ m\neq0\ \ $ & \quad$N=0$ & \qquad$E=\pm v\sqrt{m^2+p_z^2}\quad$
	  & \quad\!$\left\{\begin{array}{l}N=\frac{\Delta_0}{|\Delta_0|} \\
			   N=0\end{array}\right.$
  & \qquad\!\!$\begin{array}{ll}E=\pm vp_z\ &\ (m^2<\mu^2+\Delta_0^2) \\
	       \text{\,none} &\ (m^2>\mu^2+\Delta_0^2)\end{array}\quad$ \\
  \hline\hline
 \end{tabular}
\end{table*}

\section{Summary and concluding remarks \label{sec:summary}}
Motivated by the recent discoveries of the topological insulators and
superconductors, we have investigated the topological properties of
superconducting Dirac fermions in 3D both for the even parity pairing
and the odd parity pairing.  The results are summarized in
Table~\ref{tab:summary}.  In the chiral limit $m=0$, we find that the
system is topologically nontrivial in the sense that each of the
right-handed and left-handed sectors of the free space Hamiltonian has
the nonzero topological charge; $N_{R,L}\neq0$.  Accordingly, a vortex
line in the superconductor supports localized gapless right-handed and
left-handed fermions.  Their dispersion relations are given by
$E=\pm vp_z$, where $|v|\leq1$ defined in Eq.~(\ref{eq:velocity_even})
is the parameter dependent velocity, and thus they propagate in opposite
directions along the vortex line.

The effect of the nonzero fermion mass $m\neq0$ is qualitatively
different for the even parity pairing and the odd parity pairing.  When
the pairing gap is parity even, the presence of the fermion mass
immediately opens up a mass gap for the fermions localized around the
vortex line and the dispersion relations become
$E=\pm v\sqrt{m^2+p_z^2}$.  This can be understood from the vanishing
total topological charge of the free space Hamiltonian for the even
parity pairing; $N(=N_R+N_L)=0$.

On the other hand, when the pairing gap is parity odd, the total
topological charge of the free space Hamiltonian is nonvanishing
$N\neq0$ and thus the system remains topological even in the presence of
the small fermion mass.  Accordingly, the vortex line still supports the
localized gapless fermions.  When the fermion mass is increased further,
there is a topological phase transition at $m=\sqrt{\mu^2+\Delta^2}$,
where the topological charge jumps from the nonzero value to zero and
consequently the gapless fermions localized around the vortex line
disappear.

Our results for the even parity pairing are relevant to the color
superconductivity of quarks.  In the CFL phase where all nine quarks are
gapped, the mean-field model Hamiltonian for the color superconductor
(\ref{eq:CSC}) has the topological charge
$N_R=-N_L=\sum_{j=1}^9\frac{\Delta_j}{2|\Delta_j|}$
in the chiral limit $m=0$.  Therefore the $\mathrm{U(1)_B}$ vortex line
that arises when the CFL quark matter is
rotated~\cite{Forbes:2001gj,Iida:2002ev} supports nine sets of localized
gapless right-handed and left-handed quarks.  In the presence of the
small quark mass or chiral condensate, such localized quarks become
gapped but their mass gap $vm$ is much smaller than the energy gap of
bulk quarks $\Delta$.\footnote{Supposing $\mu\sim500$ MeV,
$\Delta\sim50$ MeV, and $m\sim10$ MeV and using
Eq.~(\ref{eq:velocity_even2}), we can estimate the mass gap to be as
small as $vm/\Delta\sim5\times10^{-3}$.}  Furthermore, the mass gap at
high density $vm\sim m\left(\Delta/\mu\right)^2\ln(\mu/\Delta)$ [see
Eq.~(\ref{eq:velocity_even2})] is parametrically smaller than the masses
of pseudo-Nambu-Goldstone bosons $\sim m\left(\Delta/\mu\right)$ in the
CFL phase~\cite{Son:1999cm}, which are important to the transport
properties and neutrino emissivity of the CFL quark
matter~\cite{Alford:2007xm}.  Whether such new low-energy degrees of
freedom localized around the vortex line have some impact on the physics
of rotating neutron/quark stars is an important problem and should be
investigated in a future work.

Our results for the odd parity pairing might be irrelevant to the color
superconductivity of quarks.  Nevertheless it would be interesting to
consider if the topological phase transition found in this paper can be
realized in condensed matter systems where the 2D Dirac fermions
appear.  

Finally, Table~\ref{tab:summary} reveals the intriguing connection
between the nonzero topological charge of the free space Hamiltonian and
the existence of localized gapless fermions in the presence of a
vortex.  We also elucidated the existence of gapless surface fermions
localized at a boundary when two phases with different topological
charges are connected.  The mathematical proof of these correspondences
remains an open question.

{\em Note added\/}.---%
After the submission of this paper, there appeared a paper by
S.~Yasui, K.~Itakura, and M.~Nitta~\cite{Yasui:2010} which has some
overlap with the present paper.  In their paper, a non-Abelian vortex in
the CFL phase is also considered.

\acknowledgments
The author thanks R.~Jackiw and F.~Wang for many useful discussions and
K.~Rajagopal for reading the manuscript.  This work was supported by MIT
Pappalardo Fellowships in Physics.

\appendix

\section{Derivations of solutions for even parity pairing
 \label{sec:derivation_even}}
Here we outline how the solutions to the Bogoliubov--de Genne equation
(\ref{eq:BdG_even}) for the even parity pairing are derived.  We
introduce notations
\begin{equation}
 \Phi(r,\theta,z) =
  \begin{pmatrix}
   F_R \\ F_L \\ G_R \\ G_L
  \end{pmatrix}
  e^{ip_zz},
\end{equation}
where $F_{R(L)}$ and $G_{R(L)}$ are right-handed (left-handed)
two-component fields.  We first make an ansatz
\begin{equation}
 F_{R}(r,\theta) =
  \begin{pmatrix}
   f_{R\uparrow}(r) \\ e^{i\theta}f_{R\downarrow}(r)
  \end{pmatrix}
  e^{-\int_0^r|\Delta(r')|dr'}
\end{equation}
and
\begin{equation}
 G_{R}(r,\theta) =
  \begin{pmatrix}
   e^{-i\theta}g_{R\uparrow}(r) \\ g_{R\downarrow}(r)
  \end{pmatrix}
  e^{-\int_0^r|\Delta(r')|dr'}
\end{equation}
and the same for $R\to L$ so that they are exponentially localized in
the $x$-$y$ plane.  Then we look for $f_{R(L)}$ and $g_{R(L)}$ that are
regular at origin and independent of $|\Delta(r)|$.  $|\Delta(r)|$ can
be eliminated from the equations by imposing
\begin{equation}
 g_R = -i\sigma_1 f_R \qquad\text{and}\qquad g_L = i\sigma_1 f_L.
\end{equation}
Now $f_R$ and $f_L$ satisfy the following four sets of equations:
\begin{subequations}
 \begin{align}
  &
  \begin{pmatrix}
   -\mu & \frac1{i}\left(\d_r+\frac{1}{r}\right) \\
   \frac1{i}\d_r & -\mu
  \end{pmatrix}
  f_R = 0 \\
  &
  \begin{pmatrix}
   p_z-E & 0 \\
   0 & -p_z-E
  \end{pmatrix}
  f_R + m f_L = 0 \\
  &
  \begin{pmatrix}
   \mu & \frac1{i}\left(\d_r+\frac{1}{r}\right) \\
   \frac1{i}\d_r & \mu
  \end{pmatrix}
  f_L = 0 \\
  &
  \begin{pmatrix}
   -p_z-E & 0 \\
   0 & p_z-E
  \end{pmatrix}
  f_L + m f_R = 0.
 \end{align}
\end{subequations}
We can find consistent solutions in two cases; $m=p_z=E=0$
[Eqs.~(\ref{eq:zero-energy_even1}) and (\ref{eq:zero-energy_even2})] and
$\mu=0$ [Eq.~(\ref{eq:nonzero_even})].

\section{Derivations of solutions for odd parity pairing
 \label{sec:derivation_odd}}
Here we outline how the solutions to the Bogoliubov--de Genne equation
(\ref{eq:BdG_odd}) for the odd parity pairing are derived.  We use the
same notations as in Appendix~\ref{sec:derivation_even}.  In this case,
$|\Delta(r)|$ can be eliminated from the equations by imposing
\begin{equation}
 g_R = -i\sigma_1 f_R \qquad\text{and}\qquad g_L = -i\sigma_1 f_L.
\end{equation}
Now $f_R$ and $f_L$ satisfy the following four sets of equations:
\begin{subequations}
 \begin{align}
  &
  \begin{pmatrix}
   -\mu & \frac1{i}\left(\d_r+\frac{1}{r}\right) \\
   \frac1{i}\d_r & -\mu
  \end{pmatrix}
  f_R + m f_L = 0 \\
  &
  \begin{pmatrix}
   p_z-E & 0 \\
   0 & -p_z-E
  \end{pmatrix}
  f_R = 0 \\
  &
  \begin{pmatrix}
   \mu & \frac1{i}\left(\d_r+\frac{1}{r}\right) \\
   \frac1{i}\d_r & \mu
  \end{pmatrix}
  f_L - m f_R = 0 \\
  &
  \begin{pmatrix}
   -p_z-E & 0 \\
   0 & p_z-E
  \end{pmatrix}
  f_L = 0.
 \end{align}
\end{subequations}
We can find consistent solutions in two cases; $p_z=E=0$
[Eqs.~(\ref{eq:zero-energy_odd1}) and (\ref{eq:zero-energy_odd2})] and
$\mu=0$ [Eqs.~(\ref{eq:nonzero_odd1}) and (\ref{eq:nonzero_odd2})].

\section{Derivations of solutions for boundary problems
 \label{sec:derivation_boundary}}
Here we outline how the spectrum of fermions in the presence of the
boundary (\ref{eq:boundary}) studied in Sec.~\ref{sec:sign_change} is
obtained.  The Bogoliubov--de Genne equation to be solved is
\begin{equation}
 \begin{pmatrix}
  -i\bm\alpha\cdot\bm\d+\beta m-\mu & (\gamma_5)\Delta(z) \\
  (\gamma_5)\Delta(z) & i\bm\alpha\cdot\bm\d-\beta m+\mu
 \end{pmatrix}
 \Phi(\x) = E \Phi(\x).
\end{equation}
Because of the translational invariance in the $x$-$y$ plane, we look
for solutions of the form
\begin{equation}
 \Phi(\x) = e^{ip_xx+ip_yy}\phi_{p_x,p_y}\!(z).
\end{equation}
We first make an ansatz
\begin{equation}\label{eq:wave_function}
 \phi_{p_x,p_y}\!(z) =
  \begin{pmatrix}
   f_R(z) \\ f_L(z) \\ g_R(z) \\ g_L(z)
  \end{pmatrix}
  e^{-\int_0^z\Delta(z')dz'}
\end{equation}
so that the solution is exponentially localized in the $z$ direction.
Then we look for $f_{R(L)}$ and $g_{R(L)}$ that are independent of
$\Delta(z)$.  $\Delta(z)$ can be eliminated from the equations by
imposing
\begin{equation}
 g_R = -i\sigma_3 f_R \qquad\text{and}\qquad g_L = \pm i\sigma_3 f_L,
\end{equation}
where the upper (lower) sign corresponds to the even (odd) parity
pairing.  Now $f_R$ and $f_L$ satisfy the following four sets of
equations:
\begin{subequations}
 \begin{align}
  &
  \begin{pmatrix}
   \frac{\d_z}{i}-\mu-E & p_- \\
   p_+ & -\frac{\d_z}{i}-\mu-E
  \end{pmatrix}
  f_R + mf_L = 0 \\
  & 
  \begin{pmatrix}
   -\frac{\d_z}{i}+\mu-E & p_- \\
   p_+ & \frac{\d_z}{i}+\mu-E
  \end{pmatrix}
  f_R \pm mf_L = 0 \\
  &
  \begin{pmatrix}
   -\frac{\d_z}{i}-\mu-E & -p_- \\
   -p_+ & \frac{\d_z}{i}-\mu-E
  \end{pmatrix}
  f_L + mf_R = 0 \\
  &
  \begin{pmatrix}
   \frac{\d_z}{i}+\mu-E & -p_- \\
   -p_+ & -\frac{\d_z}{i}+\mu-E
  \end{pmatrix}
  f_L \pm mf_R = 0
 \end{align}
\end{subequations}
with $p_\pm\equiv p_x\pm ip_y$.

\subsection{Even parity pairing}
When the pairing gap is parity even (upper sign), we can find the
following four zero energy solutions at $m=p_x=p_y=0$:
\begin{equation}
 \begin{pmatrix}
  f_R \\ f_L
 \end{pmatrix}
 \propto
\begin{pmatrix}
 e^{i\mu z} \\ 0 \\ 0 \\ 0
\end{pmatrix},
\begin{pmatrix}
 0 \\ e^{-i\mu z} \\ 0 \\ 0
\end{pmatrix},
\begin{pmatrix}
 0 \\ 0 \\ e^{-i\mu z} \\ 0
\end{pmatrix},
\begin{pmatrix}
 0 \\ 0 \\ 0 \\ e^{i\mu z}
\end{pmatrix}.
\end{equation}
By evaluating matrix elements of the perturbation Hamiltonian
\begin{equation}
 \delta\H =
  \begin{pmatrix}
   \alpha_xp_x+\alpha_yp_y+\beta m & 0 \\
   0 & -\alpha_xp_x-\alpha_yp_y-\beta m
  \end{pmatrix}
\end{equation}
with respect to the corresponding four zero energy eigenfunctions
$\phi_{0,0}(z)$ in Eq.~(\ref{eq:wave_function}), we obtain the following
effective 2D Hamiltonian:
\begin{equation}
 \H_\mathrm{2D} =
 \begin{pmatrix}
  0 & v^*p_- & v^*m & 0 \\
  vp_+ & 0 & 0 & vm \\
  vm & 0 & 0 & -vp_- \\
  0 & v^*m & -v^*p_+ & 0 \\
 \end{pmatrix}
\end{equation}
with the velocity $v$ defined in Eq.~(\ref{eq:v_boundary_even}).  Its
energy eigenvalues are given by $E=\pm|v|\sqrt{m^2+p_x^2+p_y^2}$.  This
spectrum with $v=1$ becomes exact for arbitrary $p_x$, $p_y$, and $m$ in
the special case where $\mu=0$.

\subsection{Odd parity pairing}
When the pairing gap is parity odd (lower sign), we can find the
following four zero energy solutions at $p_x=p_y=0$:
\begin{equation}\label{eq:solutions_odd}
 \begin{split}
  \begin{pmatrix}
   f_R \\ f_L
  \end{pmatrix}
  \propto &
  \begin{pmatrix}
   \mu+\sqrt{\mu^2-m^2} \\ 0 \\ m \\ 0
  \end{pmatrix}
  e^{i\sqrt{\mu^2-m^2}\,z}, \\ &
  \begin{pmatrix}
   0 \\ \mu+\sqrt{\mu^2-m^2} \\ 0 \\ m
  \end{pmatrix}
  e^{-i\sqrt{\mu^2-m^2}\,z}, \\ &
  \begin{pmatrix}
   m \\ 0 \\ \mu+\sqrt{\mu^2-m^2} \\ 0
  \end{pmatrix}
  e^{-i\sqrt{\mu^2-m^2}\,z}, \\ &
  \begin{pmatrix}
   0 \\ m \\ 0 \\ \mu+\sqrt{\mu^2-m^2}
  \end{pmatrix}
  e^{i\sqrt{\mu^2-m^2}\,z}.
 \end{split}
\end{equation}
Note that the corresponding four zero energy eigenfunctions
$\phi_{0,0}(z)$ in Eq.~(\ref{eq:wave_function}) are normalizable in the
$z$ direction as long as $m^2<\mu^2+\Delta_\infty^2$.  By evaluating
matrix elements of the perturbation Hamiltonian
\begin{equation}
 \delta\H =
  \begin{pmatrix}
   \alpha_xp_x+\alpha_yp_y & 0 \\
   0 & -\alpha_xp_x-\alpha_yp_y
  \end{pmatrix}
\end{equation}
with respect to the four zero energy eigenfunctions, we obtain the
following effective 2D Hamiltonian:
\begin{equation}
 \H_\mathrm{2D} =
 \begin{pmatrix}
  0 & v^*p_- & 0 & 0 \\
  vp_+ & 0 & 0 & 0 \\
  0 & 0 & 0 & -vp_- \\
  0 & 0 & -v^*p_+ & 0 \\
 \end{pmatrix}
\end{equation}
for $m^2<\mu^2$ with the velocity $v$ defined in
Eq.~(\ref{eq:v_boundary_odd1}) and
\begin{equation}
 \H_\mathrm{2D} =
 \begin{pmatrix}
  0 & 0 & 0 & -ivp_- \\
  0 & 0 & -ivp_+ & 0 \\
  0 & ivp_- & 0 & 0 \\
  ivp_+ & 0 & 0 & 0 \\
 \end{pmatrix}
\end{equation}
for $\mu^2<m^2<\mu^2+\Delta_\infty^2$ with the velocity $v$ defined in
Eq.~(\ref{eq:v_boundary_odd2}).  Their energy eigenvalues are given by
$E=\pm|v|\sqrt{p_x^2+p_y^2}$.  This spectrum with $v=1$ becomes
exact for arbitrary $p_x$ and $p_y$ in the special case where $\mu=0$.

\end{document}